\documentstyle[aps, twocolumn, graphics, epsfig, floats, rmp]{revtex}
\begin{document}
\title{The essence of entanglement}
\author{\v Caslav Brukner$^1$, Marek \. Zukowski$^2$ and Anton Zeilinger$^1$}
\address{$^1$Institut f\"{u}r Experimentalphysik, Universit\"at Wien, Boltzmanngasse 5, 1090 Wien,
Austria \\ $^2$Instytut Fizyki Teoretycznej i Astrofizyki
Uniwersytet Gdanski, PL-80-952 Gdansk, Poland}

\maketitle

\begin{abstract}

Entanglement, according to Erwin Schr\"odinger {\it the essence}
of quantum mechanics, is at the heart of the
Einstein-Podolsky-Rosen paradox and of the so called
quantum-nonlocality -- the fact that a local realistic explanation
of quantum mechanics is not possible as quantitatively expressed
by violation of Bell's inequalities. Even as entanglement gains
increasing importance in most quantum information processing
protocols, its conceptual foundation is still widely debated.
Among the open questions are: What is the conceptual meaning of
quantum entanglement? What are the most general constraints
imposed by local realism? Which general quantum states violate
these constraints? Developing Schr\"odinger's ideas in an
information-theoretic context we suggest that a natural
understanding of quantum entanglement results when one accepts (1)
that the amount of information per elementary system is finite and
(2) that the information in a composite system resides more in the
correlations than in properties of individuals. The quantitative
formulation of these ideas leads to a rather natural criterion of
quantum entanglement. Independently, extending Bell's original
ideas, we obtain a single general Bell inequality that summarizes
all possible constraints imposed by local realism on the
correlations for a multi-particle system. Violation of the general
Bell inequality results in an independent general criterion for
quantum entanglement. Most importantly, the two criteria agree in
essence, though the two approaches are conceptually very
different. This concurrence strongly supports the
information-theoretic interpretation of quantum entanglement and
of quantum physics in general.

\end{abstract}

\tableofcontents

\section{INTRODUCTION}

In their seminal paper in 1935 Einstein, Podolsky and Rosen (EPR)
consider quantum systems consisting of two particles such that,
while neither position nor momentum of either particle is well
defined, both the sum of their positions and the difference of
their momenta are both precisely defined. It then follows that
measurement on either position or momentum performed on, say,
particle 1 immediately implies for particle 2 a precise position
or momentum respectively even when the two particles are separated
by arbitrary distances without any actual interaction between
them.

Motivated by EPR Schr\"odinger (1935a) in his paper entitled  "The
present situation in quantum mechanics" wrote succinctly "Maximal
knowledge of a total system does not necessarily include total
knowledge of all its parts, not even when these are fully
separated from each other and at the moment are not influencing
each other at all" and he coined the term\footnote{Schr\"odinger
(1935a) used the German original "Verschr\"ankung" and he himself
introduced the English translation "entanglement" (Schr\"odinger,
1935b).} "entanglement of our knowledge" to describe this
situation. A central goal of our work is to take Schr\"odinger's
position as a starting point for a quantitative
information-theoretic definition of entanglement. Therefore,
carefully reading his sentence one may identify three independent
ideas on which Schr\"odinger builds his notion of entanglement.
(a) First, our knowledge, or total information, of a system is
bound. (b) Second, the total information of a composite system is
not necessarily fully contained in its individual constituents.
(c) And the third, these statements are independent of the
relative space-time arrangements of the individual observations on
the constituents of the composite system. Furthermore  we
underline that Schr\"odinger talks about quantum states
representing our expectation catalogs and he carefully avoids the
notion of properties of systems.

In 1964 John Bell obtained certain bounds (Bell inequalities) on
combinations of statistical correlations for measurements on
two-particle systems if these correlations are understood within a
realistic picture based on local hidden properties of each
individual particle. In a realistic picture the measurement
results are determined by properties the particles carry prior to
and independent of observation. In a local picture the results
obtained at one location are independent of any measurements or
actions performed at space-like separation. Then Bell showed that
quantum mechanics predicts violation of these constrains for
certain statistical predictions for two-particle systems. A more
striking conflict between quantum mechanical and local realistic
predictions even for perfect correlations has been discovered for
three and more particles (Greenberger {\it et al.}, 1989; 1990;
Mermin, 1990), resulting in an outright invalidation of the EPR
concepts. By now a number of experiments (Freedman and Clauser,
1972; Aspect {\it et al.}, 1981; Pan {\it et al.}, 2000) have
confirmed the quantum mechanical predictions even if the
individual particles are truly space-like separated (Weihs {\it et
al.}, 1998). However two important questions remain: (1) "What are
the most general constraints on correlations imposed by local
realism?" and (2) "Which quantum states violate these
constraints?". The latter has been solved in general only in the
case of two particles in pure states (Gisin, 1991; Gisin and Peres
1992) and for two-qubit mixed states (Horodecki {\it et al.},
1995). Only recently bounds for local realistic description of
higher-dimensional systems have been found in some simple cases
(Kaszlikowski {\it et al.}, 2000).

Our paper will now develop the ideas presented above in various
quantitative ways. Following first Schr\"odinger's ideas we will
define quantum entanglement as a feature of a composite system to
have more information contained in correlations than any classical
mixture of its individual constituents could ever have. The
essence of classical correlations is that there the joint
properties can be reduced to correlations between properties of
the individual constituents. This is an operational definition
because information is always defined through observation of
measurement results.

In parallel, following then Bell's ideas, we will obtain a single
general Bell inequality that summarizes all possible local
realistic constraints on the correlations for a multi-qubit
system, where two dichotomic observables are measured on each
individual qubit. This enables us to introduce another operational
definition of entangled states as those which violate that general
Bell inequality in a direct measurement\footnote{By an entangled
state is often meant a state which cannot be represented as a
classical mixture of product states (non-separable state). Here,
since our aim is to relate Schr\"odinger's notion of entanglement
to the one via Bell's inequality we will use the definition as
given in the text above. However, it is well known that there are
cases of "hidden non-locality" where a quantum state initially
does not directly violate a Bell inequality but after local
operations together with classical communication such a violation
might occur (Popescu, 1995).}. Finally we show that the two
operational definitions are equivalent in the two-qubit case and
that in general our informational definition of entanglement
provides a necessary and rather stringent condition for violating
the general Bell inequality. We find this intriguing because the
two operational approaches are based on completely different
concepts. To us this further supports the view that information is
the most fundamental concept in quantum physics (Zeilinger, 1999).

\section{FINITENESS OF INFORMATION}

A central point in our discussion will be the different ways how
information can be distributed within a composite system. We
therefore have to introduce our notion of information and we have
to give our considerations about how much information altogether a
system can represent.

Any physical description of a physical system is a set of
propositions together with their truth values - true or false.
Then, any proposition we might assign to a quantum system which is
always based on observation of properties of the classical
apparatus used, represents our knowledge, i.e., information, of a
system gained through observation. To illustrate this consider the
state $|\psi\rangle=|z+\rangle$ of a spin-1/2 particle with spin
up along the $z$-axis, which is an eigenstate of the operator
$\sigma_z$ with eigenvalue +1. This simply means that the quantum
system described by the state  $|\psi\rangle$ will be found with
certainty to have spin +1 if it is measured along the $z$-axis.
Thus the information content in that state is represented by the
truth value of the proposition: ''The spin along the $z$-axis is
up.'' This is one bit of information. It is clear that both truth
values must be possible. Only then observation of the system can
result in a gain of information. In agreement with Schr\"odinger's
idea (a) above the most simple system then just represents one bit
of information. This is what we mean when we talk about a system
carrying information. To us a system is a construct based on
information.

If we assume that one bit is the only information the most simple
quantum system can carry, and that this is defined with respect to
a certain measurement then other measurements must contain an
element of irreducible randomness. Otherwise the system would
carry more information in conflict with (a) for a system
representing one bit of information only. This means that there
are other measurement directions for which the experimental
outcome is completely random. Specifically, for a measurement
along any direction $\vec{n}$ in the $x$-$y$ plane the proposition
''The spin along the $\vec{n}$  axis is up'' is completely
indefinite, i.e., we have absolutely no knowledge which particular
outcome "spin up" or "spin down" will be observed in an individual
experimental trial. The two propositions about spin along $z$ and
about spin along $\vec{n}$ are propositions with a property of
mutual exclusiveness. This is quantum complementarity: the
complete knowledge of the truth value of one of the propositions
implies maximal uncertainty about the truth values of the other.

How much information is carried by a system with respect to a
specific set of mutually complementary propositions? We suggest
that it is natural to assume that the information contained in a
set of mutually complementary propositions is the sum over the
measures of information of the individual members of that
particular set (Bohr, 1958). Specifically, to obtain the total
information carried by a quantum system one summarizes over all
individual measures of information for a complete set of mutually
complementary measurements, as shown in Brukner and Zeilinger
(1999). There it was shown that the total information carried by
the composite system consisting of N qubits in a pure state to be
N bits of information.

Here we are interested in the various ways how information can be
distributed within a composite system. In particular we will
consider that part of the total information of the system which is
contained in correlations, or joint properties of its
constituents. This is also the reason why here we do not consider
complete sets of mutually complementary propositions for the
composite system but just a subset of them concerning joint
properties of its constituents.

\section{INFORMATION CONTAINED IN CORRELATIONS}

Correlations between quantum systems have assumed a very central
role in the discussions of the foundations of quantum mechanics.
We will now investigate how much information can be contained in
such correlations in order to give an information-theoretic
criterion of quantum entanglement. As it is our final goal to
compare that criterion with the one given by Bell-type
inequalities (Clauser {\it et al.}, 1969; Mermin, 1990; Belinskii
and Klyshko, 1993) where one considers correlations between spin
measurements confined on each side within one plane we restrict
our analysis to an $x$-$y$ plane locally defined for each
subsystem.

For clarity of presentation we first investigate the case of a
two-qubit system carrying therefore N=2 bits of information, i.e.
representing the truth value of two propositions. That information
contained in 2 propositions can be distributed over the 2 qubits
in various ways.

Consider first a product state e.g.
$|\psi\rangle=|+x\rangle_1|-x\rangle_2$. Here the state
$|\psi\rangle$ represents the two-bit combination true-false of
the truth values of the propositions about the spin of each
particle along the $x$-axis: (1) "The spin of particle 1 is up
along $x$" and (2) "The spin of particle 2 is up along $x$".
Instead of the second proposition describing the spin of particle
2, we could alternatively choose a proposition which describes the
result of a joint observation: (3) "The two spins are the same
along $x$." Then the state $|\psi\rangle$ represents the two-bit
combination true-false of the truth values of the propositions (1)
and (3). Note that this is also present in classical composite
systems.

Evidently, for pure product states at most one proposition with
definite truth-value can be made about joint properties because
one proposition has to be used up to define a property of one of
the two subsystems. In other words 1 bit of information defines
the correlations
\begin{equation}
I^{prod}_{corr}=1.
\label{prodcorr}
\end{equation}
In our example where $|\psi\rangle=|+x\rangle_1|-x\rangle_2$ the
correlations are fully represented by the correlations between
$x$-measurements on both sides, therefore
$I^{prod}_{corr}=1=I_{xx}=1$. The specific measure of information
used by us will be specified below.

Obviously, the choice of directions within the planes of
measurement on the two sides is arbitrary. Choosing general $x$
and $y$ directions we request that the total information contained
in the correlations must be invariant upon this choice. This
invariance property by itself already defines the specific measure
of information and it rules out Shannon's measure (Brukner and
Zeilinger, 2001). Following our arguments given above we now
define the information contained in the correlations as the sum
over the individual measures of information carried in a complete
set of mutually complementary observations within $x$-$y$ plane.
Therefore the information contained in the correlations is
quantified by the sum
\begin{equation}
I_{corr}=I_{xx}+I_{xy}+I_{yx}+I_{yy} \label{corr}
\end{equation}
of the partial measures of information contained in the set of
complementary observations within the $x$-$y$-plane.  These
observations are mutually complementary for {\it product states}
and the set is complete as there exists no further complementary
observation within the chosen planes. By this we mean that for any
product state a complete knowledge contained in one of the
observations in Eq. (\ref{corr}) excludes any knowledge content in
the other three observations.

Consider now a maximally entangled Bell state, e.g.
\begin{eqnarray}
|\phi^-\rangle&=&\frac{1}{\sqrt{2}}(|+x\rangle_1|-x\rangle_2+|-x\rangle_1|+x\rangle_2)
\nonumber
\\
&=&\frac{1}{\sqrt{2}}(|+y\rangle_1|+y\rangle_2+|-y\rangle_1|-y\rangle_2).
\label{bellstate}
\end{eqnarray}
The two propositions here both are statements about results of
joint observations (Zeilinger 1997), namely (1') "The two spins
are equal along $x$" and (2') "The two spins are equal along $y$".
Now the state represents the two-bit combination false-true of
these propositions about correlations. Note that here the 2 bits
of information are all carried by the 2 qubits in a joint way,
with no individual qubit carrying any information on its own. In
other words, as the two available bits of information are already
exhausted in defining joint properties, no further possibility
exists to also encode information in individuals. Therefore
\begin{equation}
I^{Bell}_{corr}=2. \label{bellcorr}
\end{equation}
Note that in our example $I_{xx}=I_{yy}=1$  and $I_{xy}=I_{yx}=0$.
Also, note that the truth value for another proposition, namely,
"The two spins are equal along $z$" must follow immediately from
the truth values of the propositions (1') and (2'), as only 2 bits
of information are available. Interestingly this is also a direct
consequence of the formalism of quantum mechanics as the joint
eigenstate of   $\sigma^1_x \sigma^2_x$ and of   $\sigma^1_y
\sigma^2_y$  is also the eigenstate of $\sigma^1_z \sigma^2_z =
-(\sigma^1_x \sigma^2_x) (\sigma^1_y \sigma^2_y)$.

In contrast to product states we suggest entanglement of two
qubits to be defined in general such that {\it more than one bit}
(of the two available ones) is used to define joint properties,
i.e.
\begin{equation}
I^{entgld}_{corr}>1 \label{entanglcorr}
\end{equation}
for at least one choice of the local $x$ and $y$ directions for
the two qubits. This is in agreement with Schr\"odinger's idea
(b). Equivalently we suggest to define the two qubits as
classically composed (non-entangled) if less than or equal to one
bit of information is used to define correlations, i.e.
\begin{equation}
I^{nonent}_{corr} \leq 1 \label{nonentanglcorr}
\end{equation}
for all possible choices of the local $x$ and $y$ directions for
two qubits. As we will show below, the independent
information-theoretic definition of entanglement
(\ref{entanglcorr}) will turn out to be equivalent to a necessary
and sufficient condition for a violation of any Bell-type
inequality for two-qubits.

In the generalization to more and more qubits we consider, without
loss of generality, a product state $|\psi\rangle=
|+x_1\rangle|+x_2\rangle...|+x_N\rangle$ of N qubits. Here, $x_j$
denotes a spatial direction in a local coordinate system of
observer $j$. Then only one proposition with definite truth-value
can be made about the correlations in the N qubits, namely the
proposition (*): "The product of spin of particle 1 along $x_1$,
spin of particle 2 along $x_2$, ... and spin of particle N along
$x_N$ is +1". This means that for a product state or a classical
mixture of product states again at most one bit is represented in
N-qubit correlations. We therefore here too suggest to define N
qubits as classically composed if not more than one bit of
information is used to define correlations, i.e.
\begin{equation}
I^{nonent}_{corr}= \sum_{x_1,...,x_N \atop =x,y} I_{x_1...x_N}
\leq 1 \label{sumpartial}
\end{equation}
for all possible choices of local $x$ and $y$ directions for N
qubits. Here the sum is over the measures of information over a
set of propositions of the type given above where $x_1,...,x_N \in
\{x,y\}$ and which therefore are mutually complementary for
product states\footnote{Since the set of propositions entering Eq.
(\ref{sumpartial}) contains also propositions which are not
mutually complementary for a non-product state the sum in
(\ref{sumpartial}) does not give the value of information in bits
contained in correlations for such states. It is then not
surprising that, for example, in the case of a three-qubit GHZ
state the sum results in 4 $(I_{xyy}=I_{yxy}=I_{yyx}=I_{xxx}=1$,
other are zero). Here however only three of the four propositions
are independent, as only 3 bits of information are available. This
again is also a direct consequence of the formalism of quantum
mechanics as the joint eigenstate of $\sigma^1_x \sigma^2_y
\sigma^3_y$, $\sigma^1_y \sigma^2_x \sigma^3_y$ and $\sigma^1_y
\sigma^2_y \sigma^3_x$ is also an eigenstate of $\sigma^1_x
\sigma^2_x\sigma^3_x = -(\sigma^1_x \sigma^2_y \sigma^3_y)
(\sigma^1_y \sigma^2_x \sigma^3_y) (\sigma^1_y\sigma^2_y
\sigma^3_x).$ Nevertheless when the sum in (\ref{sumpartial}) is
larger than unity, this indicates that the total information
contained in correlations is larger than one bit.}.

We would like to make a very general comment. The quantitative
condition (\ref{sumpartial}) while certainly correct for the
situations discussed here might have to be modified in order to
apply to more complicated cases like entanglement between many
qubits when the measurements are not restricted to one plane or
entanglement between systems defined in Hilbert spaces of higher
dimensions, so-called qunits. Another interesting case can arise
when one also considers in detail all possible sets of
correlations between all possible sets of subsystems.  For example
for 3-qubit systems we have one correlation between all three
individual qubits, we have 3 correlations between two individual
qubits and we have another 3 correlations between one qubit and
the other two. This results in a large number of conditions of the
type of Eq. (\ref{sumpartial}) which are not independent from each
other in general. Yet, we stress, that it is to be expected that
the general ideas laid out here will still be applicable. A most
important guidance for quantitative conditions being that the
information carried by the correlations between subsystems exceeds
the limit given by the information carried by the subsystems
themselves.

Finally we stress that in our information-theoretic analysis of
entanglement we did not have to use concepts like spatial
separation between subsystems of the composite system or the
relative times of the observations on the subsystems. This is in
agreement with Schr\"odinger's idea (c).

To this point we did not yet specify any particular measure of
information. Thus the question arises which particular measure of
information is adequate to define the information gain in an
individual quantum experiment.

\section{QUANTIFYING INFORMATION}

Consider an experiment with two outcomes "yes" and "no" and with
the probabilities $p_1$ and $p_2=1-p_1$, respectively for the two
outcomes. Within finite time the experimenter can perform only a
finite number of experimental trials. Because of inherent
fluctuations associated with any probabilistic experiment with a
finite number of trials the number of occurrences of a specific
outcome in future repetitions of the experiment is not precisely
predictable. Rather it obeys the binomial distribution (See e.g.
Gnedenko, 1976).

If one bets for example that the number of "yes" outcomes will be
the one with highest probability, the probability of success still
depends on $p_1$. With a probability of $p_1=0.5$, the probability
of 5 "yes" outcomes in 10 trials is only 0.25, but with $p_1=0.9$
the probability of 9 "yes" outcomes in 10 trials is 0.39.  It is a
trait of the binomial distribution, that we know the future number
of occurrences of the outcomes very well if $p_1$ (or equivalently
$p_2$) is close to 0 or 1, but we know much less about them when
$p_1$ is around 0.5. Note that this follows from elementary
probability theory without any input from physics\footnote{Here a
very subtle position was assumed by von Weizs\"acker (1975) who
wrote: "It is most important to see that this [the fact that
probability is not a prediction of the precise value of the
relative frequency] is not a particular weakness of the objective
empirical use of the concept of probability, but a feature of the
objective empirical use of any quantitative concept. If you
predict that some physical quantity, say a temperature, will have
a certain value when measured, this prediction also means its
expection value within a statistical ensemble of measurements. The
same statement applies to the empirical quantity called relative
frequency. But here are two differences which are connected to
each other. The first difference: In other empirical quantities
the dispersion of the distribution is in most cases an independent
empirical property of the distribution and can be altered by more
precise measurements of other devices; in probability the
dispersion is derived from the theory itself and depends on the
absolute number of cases. The second difference: In other
empirical quantities the discussion of their statistical
distributions is done by another theory than the one to which they
individually belong, namely by the general theory of probability;
in probability this discussion evidently belongs to the theory of
this quantity, namely of probability itself. The second difference
explains the first one."}. In (Brukner and Zeilinger 1999) it was
shown that this knowledge is properly represented by the measure
\begin{equation}
I=(p_1-p_2)^2 \label{measure}.
\end{equation}
This attains its maximal value of unity when one of probabilities
is one, and it attains its minimal value of 0 when both
probabilities are equal.

Note that all our propositions about joint properties are binary
propositions, i.e. they are associated to experiments with two
possible outcomes, one of them being "the product of spin of
particle 1 along $x_1$, spin of particle 2 along $x_2$, ... and
spin of particle N along $x_N$ is +1", and the other one "the
product of spin of particle 1 along $x_1$, spin of particle 2
along $x_2$, ... and spin of particle N  along $x_N$ is -1", so
that Eq. (\ref{measure}) can be applied. If we now denote the
probabilities for the two outcomes by $p^{+}_{x_1...x_N}$ and
$p^{-}_{x_1...x_N}$ respectively, then the information contained
in proposition (*) is given by
\begin{equation}
I_{x_1...x_N}=(p^{+}_{x_1...x_N}- p^{-}_{x_1...x_N})^2.
\label{measure2}
\end{equation}

Now we will express Eq. (\ref{measure2}) in terms of the density
matrix $\rho$ of N qubits. First note that an arbitrary mixed
state of N qubits can be written as
\begin{equation}
\rho=\frac{1}{2^N} \sum_{x_1,...,x_N=0}^{3} T_{x_1...x_N} \mbox{ }
\sigma^1_{x_1} \otimes ... \otimes \sigma^N_{x_N} \label{state}
\end{equation}
where $\sigma^{j}_0$ is the identity operator in the Hilbert space
of particle $j$, and $\sigma^{j}_{x_j}$ is a Pauli operator for
$x_j=1,2,3$. Here the elements of the correlation tensor $T$ are
given as mean values of the product of the N spins,
\begin{equation}
T_{x_1...x_N}= \mbox{Tr}[ \rho (\sigma^1_{x_1} \otimes ... \otimes
\sigma^N_{x_N})]= p^+_{x_1...x_N}- p^-_{x_1...x_N}
\label{corrtensor}
\end{equation}
with $T_{0...0}=1$. Then obviously our measure of information
(\ref{measure2}) is equal to the square of the corresponding
element of the correlation tensor
\begin{equation}
I_{x_1...x_N}= T^2_{x_1...x_N}. \label{srce}
\end{equation}

We have thus obtained a quantitative expression for the individual
measures of information contained in the sum of Eq.
(\ref{sumpartial}) and we want to emphasize that our analysis of
entanglement would not be possible without the use of the measure
of information (\ref{measure}).

So far in the present paper we followed Schr\"odinger's concepts
of entangled states in our information-theoretic analysis. Now we
will follow Bell's ideas in a second, independent approach to
characterize entanglement.

\section{ALL BELL INEQUALITIES FOR CORRELATIONS}

Here we obtain a single general Bell inequality that summarizes
all possible constraints on the statistical correlations of an
N-qubit system. These constraints are derived under the
assumptions of local realism. We consider such correlation
measurements where for each individual particle one of the two
arbitrary dichotomic observables can be chosen. From this
inequality we obtain as specific corollaries the
Clauser-Horne-Shimony-Holt (CHSH) inequality (Clauser {\it et
al.}, 1969) for two-qubit systems and the related inequalities for
N qubits (Mermin, 1990; Ardehali, 1992; Belinskii and Klyshko,
1993).

In a local realistic picture one assumes that the result of every
measurement of an observable is predetermined.
Thus in such a picture one implicitly requires an unlimited amount
of information to be carried by an individual particle, which
conflicts with Schr\"odinger's idea (a). Take therefore an
individual observer and allow him or her to be able to choose
between two dichotomic observables (determined by some parameters
denoted here $\vec{n}_1$ and $\vec{n}_2$). This implies the
existence of two numbers $A_j(\vec{n}_1)$ and $A_j(\vec{n}_2)$
each taking values +1 or -1 which describe the predetermined
result of a measurement by the observer of the observable defined
by the local parameter $\vec{n}_1$  and $\vec{n}_2$, respectively.
We choose such a notation for brevity; of course each observer can
choose independently two arbitrary directions.

In a specific run of the experiment the correlations between all N
observations can be represented by the product  $\prod_{j=1}^N
A_j(\vec{n}_{k_j})$, with $k_j=1,2$. The correlation function is
then the average over many runs of the experiment
\begin{equation}
E(\vec{n}_{k_1},...,\vec{n}_{k_N})=\left\langle  \prod_{j=1}^N
A_j(\vec{n}_{k_j})\right\rangle_{avg}. \label{corrfunction}
\end{equation}
Note that for each observer $j$ one has either
$|A_j(\vec{n}_1)+A_j(\vec{n}_2)|=0$ and
$|A_j(\vec{n}_1)-A_j(\vec{n}_2)|=2$ or the other way around. Then
for all sign sequences of $s_1,...,s_N$, where $s_j\in\{-1,1\}$
the modulus of the product $|\prod_{j=1}^N [A_j(\vec{n}_{1}) + s_j
A_j(\vec{n}_{2})]|$ vanishes except just one for which the product
is $2^N$. Therefore one has
\begin{equation}
\sum_{s_1,...,s_N \atop = -1,1}  |\prod_{j=1}^N [s_j
A_j(\vec{n}_{1}) + s^2_j A_j(\vec{n}_{2})]|=2^N.
\end{equation}

It then follows directly that the correlation functions must
satisfy the following general Bell inequality (Weinfurter and
\.{Z}ukowski, 2001; \.{Z}ukowski and Brukner, 2001; for an
independent derivation see Werner and Wolf, 2001)
\begin{equation}
\sum_{s_1,...,s_N \atop =-1,1} |\sum_{k_1,...,k_N \atop = 1,2}
s^{k_1}_1 ... s^{k_N}_N \mbox{ }
E(\vec{n}_{k_1},...,\vec{n}_{k_N})|\leq 2^N. \label{thebellineq}
\end{equation}
Therefore within each modulus we have sums of all $2^N$
correlation functions, which are the result of the Bell-type
experiment, however each correlation functions is multiplied by a
specific sign  $\pm$. Using the generalized triangle inequality we
obtain the set of Bell-type inequalities which are equivalent to
inequality (\ref{thebellineq})
\begin{equation}
|\hspace{-1mm} \sum_{s_1,...,s_N \atop = -1,1} \hspace{-2mm}
S(s_1,...,s_N) \hspace{-2mm} \sum_{k_1,...,k_N \atop = 1,2}
s^{k_1}_1 ... s^{k_N}_N E(\vec{n}_{k_1},...,\vec{n}_{k_N})| \leq
2^N, \label{allbellineq}
\end{equation}
where $S(s_1,...,s_N)$ is one of the $2^{2^N}$ sign functions of
$s_1,...,s_N$, by which we mean that its possible values can only
be +1 or -1. Actually, inequalities (\ref{allbellineq}) represent
the complete set\footnote{For an extensive classification of the
inequalities see Werner and Wolf (2001). For three qubits a
complete set of inequalities has been found numerically by
Pitowsky and Svozil (2000). See also Pitowsky (1989) and Peres
(1999).} of all possible $2^{2^N}$ inequalities for the
correlations, one for each possible choice of the sign function S.
Many of these inequalities are trivial (for example when the
choice is $S(s_1,...,s_N)=1$ for all arguments, we get that the
modulus of the correlation function does not exceed 1).

An exemplary stringent condition result for $S(s_1,...,s_N)
=\sqrt{2} \cos(-\frac{\pi}{4}+(s_1+...+s_N-N)\frac{\pi}{2})$ which
leads to the series of inequalities derived by Belinskii and
Klyshko (1993). Specifically, for $N=2$, the CHSH inequality
\begin{equation}
|E(1,1)+E(1,2)+E(2,1)-E(2,2)|\leq 2 \label{CHSH}
\end{equation}
follows. For $N=3$, one obtains
\begin{equation}
|E(1,2,2)+E(2,1,2)+E(2,2,1)-E(1,1,1)| \leq 2, \label{mermin}
\end{equation}
where here we use numbers $1$ and $2$ to denote directions
$\vec{n}_1$ and $\vec{n}_2$, respectively. Inequality
(\ref{mermin}) leads to the Greenberger-Horne-Zeilinger
(Greenberger {\it et al.} 1989; 1990) contradiction for an
appropriate choice of local settings. In these cases the left hand
side of inequality (\ref{mermin}) reaches the value 4 which is the
maximum possible value for any, not only quantum, correlation
function.

Thus far we have shown that when a local realistic model applies,
the general Bell inequality (\ref{thebellineq}) follows. The
reverse is also true and we give the proof below: whenever
inequality (\ref{thebellineq}) holds one can construct a local
realistic model for the correlation function. This establishes the
general Bell inequality presented above as a necessary and
sufficient condition for local realistic description of
multi-particle correlations, where two dichotomic observables are
measured on each individual particle.

The proof of the sufficiency of condition (\ref{thebellineq}) will
be done here in a constructive way. Simply one ascribes to the set
of predetermined local results, which satisfy the following
conditions $A_j(\vec{n}_1)=s_jA_j(\vec{n}_2)$, the hidden
probability $p(s_1,...,s_N)=\frac{1}{2^N}| \sum_{k_1,...,k_N}
s^{k_1}_1 ... s^{k_N}_N E(\vec{n}_{k_1},...,\vec{n}_{k_N})|$, and
one demands that the product $\prod_{j=1}^N A_j(\vec{n}_{2})$ has
the same sign as that of the expression inside of the modulus
defining the $p(s_1,...,s_N)$. In this way a definite set of local
realistic values is ascribed a unique global hidden probability.
Obviously, such defined probabilities are positive. However due to
inequality (\ref{thebellineq}) they may add up to less than 1. In
such a case, the "missing" probability is ascribed to an arbitrary
model of local realistic noise (e.g., for which all possible
products of local results enter with equal weights). The overall
contribution of such a noise term to the correlation function is
zero. In this way we obtain a local realistic model of a certain
correlation function. However, one should check that this
construction indeed reproduces the model for the correlation
function for the set of settings that enter inequality
(\ref{thebellineq}), that is for
$E(\vec{n}_{k_1},...,\vec{n}_{k_N})$. For simplicity take $N=2$.
Notice that the expansion coefficients of the four-dimensional
vector $(E(\vec{n}_1,\vec{n}_1),E(\vec{n}_1,\vec{n}_2),
E(\vec{n}_2,\vec{n}_1),E(\vec{n}_2,\vec{n}_2))$ in terms of
orthogonal basis vectors $(s_1s_2,s^2_1s_2,s_1s^2_2, s^2_1s^2_2)$
(recall that $s_1,s_2 \in \{-1,1\}$) are equal to the expressions
within the moduli entering inequality (\ref{thebellineq}). Next
notice, that by the construction shown above the local realistic
model for $N=2$ gives
$(E_{LV}(\vec{n}_1,\vec{n}_1),E_{LV}(\vec{n}_1,\vec{n}_2),
E_{LV}(\vec{n}_2,\vec{n}_1),E_{LV}(\vec{n}_2,\vec{n}_2))$
$=\frac{1}{4} \sum_{s_1,s_2} \sum_{k_1,k_2} s^{k_1}_1 s^{k_2}_2
E(\vec{n}_{k_1},\vec{n}_{k_2}) (s_1s_2,s^2_1 s_2,s_1 s^2_2, s^2_1
s^2_2)$. Thus, since the vector built out of the correlation
function values and its local realistic counterpart have the same
expansion coefficients, they are equal and the sufficiency of
(\ref{thebellineq}) as a condition for local realism is proven.
The generalization to an arbitrary N is obvious.

Above we derived the full set of Bell inequalities for multi-qubit
correlations. This strictly defines the boundary of the validity
of local realism. We will now discuss states which violate such
inequalities with the specific aim of investigating the way
information can be distributed between the subsystems of such
states.

\section{N QUBITS THAT VIOLATE LOCAL REALISM}

Let us consider the general N-qubit state as in Eq. (\ref{state}).
Then the N-qubit quantum correlation function for a Bell-GHZ type
experiment is
\begin{eqnarray}
E_{QM}(\vec{n}_{k_1},...,\vec{n}_{k_N})&=&\mbox{Tr}[\rho(\vec{\sigma}\cdot
\vec{n}_{k_1}  \otimes ... \otimes  \vec{\sigma}\cdot
\vec{n}_{k_N})]  \label{corrquantum} \\ &=&\hspace{-5mm}
\sum_{x_1,...,x_n=1}^3 \hspace{-3mm} T_{x_1...x_N}
(\vec{n}_{k_1})_{x_1} ... (\vec{n}_{k_N})_{x_N}
\end{eqnarray}
where $(\vec{n}_{k_j})_{x_j}$ $(x_j=1,2,3)$ are the three
Cartesian components of the vector $\vec{n}_{k_j}$. Equation
(\ref{corrquantum}) means that the N-particle correlation function
is fully defined by a tensor $\hat{T}$ (the indices of which can
take values 1,2,3, and which belongs to $R^{3N}$). For convenience
we shall write down the last equation in a more compact way as
$E_{QM}(\vec{n}_{k_1},...,\vec{n}_{k_N})=\langle\hat{T},\vec{n}_{k_1}
\otimes ... \otimes \vec{n}_{k_N}\rangle$, where $\langle
...,...\rangle$ denotes the scalar product in $R^{3N}$.

The necessary and sufficient condition (\ref{thebellineq}) for a
local realistic description of  N-particle correlations implies
that the quantum correlations for N qubits can always have a local
and realistic model for the Bell-type experiment  if and only if
\begin{equation}
\sum_{s_1,...,s_N \atop = -1,1} |\langle\hat{T},\sum_{k_1=1}^2
s^{k_1}_{1} \vec{n}_{k_1} \otimes ... \otimes \sum_{k_N=1}^2
s^{k_N}_N \vec{n}_{k_N}\rangle| \leq 2^N \label{statojaradim}
\end{equation}
for any possible choice $\vec{n}_{k_1},...,\vec{n}_{k_N}$ of each
observer's two local settings $\vec{n}_1$ and $\vec{n}_2$.

This condition can be simplified further, provided one notices
that for each observer there always exist two mutually orthogonal
unit vectors $\vec{a}_1$ and $\vec{a}_2$, independently defined
for each observer, and the angle $\alpha_j$ such that
$\sum_{k_j=1}^2 \vec{n}_{k_j} = 2 \vec{a}_1 \cos(\alpha_j +
\frac{\pi}{2})$ and $\sum_{k_j=1}^2 (-1)^{k_j} \vec{n}_{k_j} = 2
\vec{a}_2 \cos(\alpha_j + \pi)$. Denoting with $c_{x_j} =
\cos(\alpha_j + x_j \frac{\pi}{2})$ one can write the inequality
(\ref{statojaradim}) as
\begin{equation}
\sum_{x_1,...,x_N \atop =1,2} |c_{x_1} ... c_{x_N} \langle
\hat{T}, \vec{a}_{x_1} \otimes ... \otimes \vec{a}_{x_N}\rangle|
\leq 1.
\end{equation}
One can rewrite this inequality as
\begin{equation}
\sum_{x_1,...,x_N \atop = 1,2} |c_{x_1} ... c_{x_N} \langle
\hat{T}, \vec{a}_{x_1} \otimes ... \otimes \vec{a}_{x_N}\rangle|
\leq 1 \label{necsufcondition}
\end{equation}
where $T_{x_1...x_N}$ is now a component of the tensor $\hat{T}$
in a new set of local coordinate systems, which among their basis
vectors have $\vec{a}_1$ and $\vec{a}_2$ which serve as the unite
vectors which define the directions $x$ and $y$.

The necessary and sufficient condition for impossibility of any
local realistic description of N-qubit correlations is that the
maximum of the left-hand side of inequality
(\ref{necsufcondition}) is larger than one. Once the values of the
elements of the correlation tensor are given for the specific
density matrix one can check via maximization procedure whether
the local realistic description is possible.

On the other hand, it is easy to notice that with the aid of the
Cauchy-Schwarz  inequality, one has an inequality
\begin{equation}
\sum_{x_1,...,x_N \atop = 1,2} |c_{x_1} ... c_{x_N} T_{x_1...x_N}|
\leq \sqrt{\sum_{x_1,...,x_2 \atop =1,2} T^2_{x_1...x_2}}.
\label{cauchyschwarz}
\end{equation}
Combining inequality (\ref{necsufcondition}) with
(\ref{cauchyschwarz}) we obtain that a sufficient condition for
the possibility of a local realistic description of the quantum
N-qubit correlations in any Bell-type experiment is that
\begin{equation}
\sum_{x_1,...,x_N \atop = 1,2} T^2_{x_1...x_N} \leq 1 \label{yeah}
\end{equation}
for all possible choices of local coordinate systems for N qubits
as then the full set of Bell inequalities (\ref{allbellineq}) is
satisfied. Equivalently, if at least one of the Bell inequalities
from the set is violated then condition (\ref{yeah}) is violated
for at least one choice of local coordinate systems.

Our measure of information (\ref{srce}) is exactly equal to the
square of the corresponding element of the quantum correlation
matrix. This establishes the equivalence between the condition for
local realism (\ref{yeah}) on correlation tensor elements and our
information-theoretical criterion (\ref{sumpartial}) for
information contained in correlations by states which do not
reveal quantum entanglement. If the state violates at least one of
the Bell inequalities from the full set (\ref{allbellineq}) then
this state is characterized by the information-theoretical
criterion (\ref{entanglcorr}).

By performing rotations in the $x$-$y$ planes of the N observers
one can vary the values of the elements of the correlation tensor,
but these variations do not change the left-hand side of
inequality (\ref{yeah}). In information-theoretic language we say
that the total information content in $x$-$y$ plane correlations
is invariant under these variations. The invariance property
implies that one can find local coordinate systems for which some
of the correlation tensor elements vanish thus having criterion
(\ref{yeah}) which involves a smaller number of them (For the case
of three qubits see Scarani and Gisin, 2001). For example, in the
two-qubit case the rotations in the $x$-$y$ planes of the two
observers are obtained with the use of two parameters, each
describing the rotation angle for the given local observer, and
therefore one can always find local coordinate systems such that
two of correlation tensor elements vanish $(T_{xy}=T_{yx}=0)$ (See
Horodecki and Horodecki, 1996). Then it can easily be seen that
varying the two angles $\alpha_1$ and $\alpha_2$ the expression on
the left hand side of inequality (\ref{cauchyschwarz}) can be
saturated by the one on the right hand side. Finally this
establishes the condition (\ref{yeah}) for two qubits as the
necessary and sufficient condition for the correlations measured
on an arbitrary two-qubit mixed state to be understood within the
local realistic picture. In that case our necessary and sufficient
condition $I_{xx}+I_{yy}>1$ (or equivalently
$T^2_{xx}+T^2_{yy}>1$) for violation of the most general Bell's
inequality is equivalent with the necessarily and sufficient
condition for violation of the CHSH inequality which was obtained
by the Horodeckis (1995). Thus, our result also confirms that
non-violation of the CHSH inequality is a necessary and sufficient
condition for the local realistic description of two-qubit
correlations.

We will now analyze from our information theoretic point the case
of an N-qubit Bell-type experiment. We are specifically interested
in the limit up to which the experiment still has a local
realistic interpretation. Consider the state which is a mixture of
the maximally entangled state and the noise induced by
experimental imperfections. Such a state is known as the Werner
state and has the form
\begin{equation}
\rho_W = V |\psi_{GHZ} \rangle \langle \psi_{GHZ}| + (1-V)
\rho_{noise}
\end{equation}
where
$ |\psi_{GHZ} \rangle = \frac{1}{\sqrt{2}} (|+z
\rangle_1...|+z\rangle_N + |-z\rangle_1...|-z\rangle_N)$
is the maximal entangled (GHZ) state and $\rho_{noise} =
\frac{1}{2^N}I$ is the completely mixed state. Here e.g.,
$|+z\rangle_j$ denotes the spin up of particle $j$ along $z$. We
would like to emphasize that the weight $V$ of the GHZ-state can
operationally be interpreted as the visibility observed in a
multi-particle interference experiment (Belinskii and Klyshko,
1993).

For the purposes of our argument we will now calculate the number
of non-zero correlation tensor elements (which are related to our
individual measures of information contained in the correlations)
for the Werner state. Note first that for any measurement
direction $\vec{n}$ belonging to the $x$-$y$ plane the spin
component $\vec{n} \cdot \vec{\sigma}$ has its eigenvectors in the
form $|\pm \vec{n}\rangle =\frac{1}{\sqrt{2}}(|+z\rangle \pm e^{i
\phi} |-z\rangle)$, where $\phi$ is the azimuthal angle of the
vector $\vec{n}$. Using Eq. (\ref{corrquantum}) one can easily
show that the correlation function for arbitrary chosen
measurement directions within the local $x$-$y$ plains is
\begin{equation}
E_{W}(\phi_1,...,\phi_N)= V \cos(\sum^N_{k=1} \phi_k).
\end{equation}
This implies that the correlation tensor elements $T_{x_1...x_N}$
with $x_1,...,x_N$ each being either $x$ or $y$ are given by
$T_{x_1...x_N}= V \cos(m_y \frac{\pi}{2})$, where $m_y$ is the
number of $y$'s in $\{x_1,...,x_N\}$. Therefore, for each N one
always has $T_{xx..x}=V$. The other components are zero except for
those that contain an even number of $y$'s, which are either V or
-V. This results in the total number $1+\sum_{k=1}^{N/2} {N
\choose 2k} = 2^{N-1}$ of non-zero components for the even N and
$1+\sum_{k=1}^{(N-1)/2} {N \choose 2k} = 2^{N-1}$ for the odd N.

We would like to stress again the equivalence
$I_{x_1...x_N}=T^2_{x_1...x_N}$ between the measure of information
contained in the correlation between measurements performed along
directions $x_1,...,x_N$ and the square of the corresponding
correlation tensor element. In the case of N qubits in the Werner
state we therefore have $2^{N-1}$ individual measures of
information with the value $V^2$ and we have the remaining ones
equal to zero. Inserting these values into the information
criterion (\ref{sumpartial}) (or equivalently (\ref{yeah})) one
obtains
\begin{equation}
V \leq \left(\frac{1}{\sqrt{2}}\right)^{N-1} \label{frka}
\end{equation}
for the maximal visibility which still allows the correlations
between N qubits in the Werner state to be understood within a
local realistic picture. Note that in such a case the right-hand
side of the inequality (\ref{cauchyschwarz}) is one. Since the
value given on the right-hand side of (\ref{frka}) is also the
minimal visibility necessarily to violate the inequalities for N
qubits obtained by Belinskii and Klyshko (1993) and since they are
included in our set of all possible inequalities
(\ref{allbellineq}) we therefore can conclude that our information
criterion (\ref{sumpartial}) is the necessary and sufficient
condition for correlations between N qubits in the Werner state to
violate the local realistic description.

It is interesting to note that different inequalities can be
obtained for correlations to be understood within local realism
where not only two but three ({\. Z}ukowski and Kaszlikowski,
1997) or even all possible measurement settings ({\. Z}ukowski,
1993) are chosen by N observers. There even lower thresholds for
the visibilities were obtained to violate the local realistic
description. For such an experimental situation our criterion
(\ref{sumpartial}) is the sufficient condition for violation of
the inequalities.

Another interesting observation is that for N maximally entangled
qubits N bits of information rest in the correlations, as opposed
to always not more than one bit for the classically composed ones.
In the case of GHZ (for $V=1$ in the consideration given above)
the information criterion (\ref{sumpartial}) results in $2^{N-1}
\leq 1$ which clearly shows that with growing N the discrepancy
between quantum and classical correlations grows exponentially.
This is in concurrence with the fact that the GHZ theorem is
stronger than Bell's and its strength, as measured by the
magnitude of violation of (\ref{allbellineq}) for maximally
entangled states, exponentially increases with the number of
qubits (Mermin, 1990; Ardehali, 1992, Belinskii and Klyshko,
1993).

\section{CONCLUSIONS}

We now would like to review what we have done in the present paper
and put it in a broad perspective. The paper contains two
independent main approaches to the question of quantum
entanglement, and we finally show their essential equivalence.

In the first approach we start from the conceptual position that
quantum mechanics is about information. We express the information
contained in composite systems such that it can be divided into
the information carried by the individuals versus the information
contained in the correlations between observations made on the
individuals. We further assume that the information contained in
any system, be it individual or composite, is finite.

Considering first the classically composed systems we note that
any correlations we might observe between the subsystems of such a
composite system can simply be understood on the basis of
correlations between the properties the individual subsystems have
on their own. This means that if we know all properties of the
individual subsystems we can definitely conclude how much
information is contained in their correlations. For the quantum
entangled systems this is not true anymore. Such composite quantum
systems can carry more information in joint properties than what
may be concluded from knowledge of the individuals. These
considerations lead to a natural information-based understanding
of quantum entanglement. Within this view we see Mermin's (1998)
"correlations without correlata" as reflecting that when
correlations are defined there is no information left to define
''correlata'' as well. "Correlations have physical meaning; that
which they correlate does not," as stated by Mermin (1998).

In an independent approach we obtain the most general set of Bell
inequalities for N qubits. That way we arrive at a necessary and
sufficient condition for quantum states whose correlations cannot
be understood within local realism. Local realism is based on the
assumption that results of the observations on the individual
systems are predetermined and independent of whatever measurements
might be performed distantly. One may notice that this assumption
implicitly says that correlations between subsystems do not go
beyond what might be concluded from the properties of individual
subsystems.

We finally show that the two approaches, the information
theoretical one and the one via Bell's inequalities, are
equivalent in their essence. This is done via the fact that the
Bell inequality criteria can be translated into a statement about
correlations (probabilities), which again can be understood as an
information theoretical expression. This requires the use of a new
measure of information introduced earlier (Brukner and Zeilinger,
1999). This measure of information is distinct from Shannon's
measure (Shannon, 1948). The main conceptual difference is that
Shannon's measure tacitly assumes that the properties of the
systems carrying the information are already well defined prior
to, and independent of, observation (Brukner and Zeilinger 2001).
In quantum mechanics this clearly is not the case. There, the
criterion for choosing the new measure of information was that it
is invariant on the experimentalist's free choice of a complete
set of mutually complementary observables.

Summing up, we would like to draw the reader's attention to the
fact of the equivalence of the two approaches in the present
paper, the information theoretic one and the one via Bell's
inequalities. It is evident that the first one is both
conceptually and formally much simpler. It is suggestive that this
new information theoretic formulation of quantum phenomena opens
up the avenue of new approaches to well known problems in quantum
information physics and in the foundations of quantum mechanics.

\section*{ACKNOWLEDGEMENTS} The work was supported by the Austrian
Science Foundation (FWF), project F1506, and by the QIPC Program
of the European Union. It is also a part of the Austrian-Polish
collaboration programme 2400. M.\.{Z}. acknowledges KBN grant No.
5 P03B 088 20.


\begin{references}

\bibitem{}
Ardehali, M., 1992, "Bell inequalities with a magnitude of
violation that grows exponentially with the number of particles,"
Phys. Rev. A {\bf 46}, 5375-5378.
\bibitem{}
Aspect, A., P. Grangier, and G. Roger, 1981, "Experimental tests
of realistic local theories via Bell's theorem," Phys. Rev. Lett.
{\bf 47}, 460-463.
\bibitem{}
Belinskii, A. V. and D. N. Klyshko, 1993, Phys. Usp. {\bf 36}, 653
(1993).
\bibitem{}
Bell, J. S., 1964, "On the Einstein-Podolsky-Rosen paradox,"
Physics 1, 195-200; reprinted Bell, J. S., 1987, {\it Speakable
and Unspeakable in Quantum Mechanics} (Cambridge Univ. Press,
Cambridge).
\bibitem{}
Bohr, N., 1958, {\it Atomic Physics and Human Knowledge} (Wiley,
New York).
\bibitem{}
Brukner, \v C., and A. Zeilinger, 1999, "Operationally invariant
information in quantum measurements," Phys. Rev. Lett. {\bf 83},
3354-3357.
\bibitem{}
Brukner, \v C. and A. Zeilinger, 2001, "Conceptual
inadequacy of the Shannon information in quantum measurements,"
Phys. Rev. A {\bf 63}, 022113 1-10.
\bibitem{}
Clauser, J., M. Horne, A. Shimony, and R. Holt, 1969, "Proposed
experiment to test local hidden-variable theories," Phys. Rev.
Lett. {\bf 23}, 880-884.
\bibitem{}
Einstein, A., B. Podolsky, and N. Rosen, 1935, "Can
quantum-mechanical description of physical reality be considered
complete?," Phys. Rev. {\bf 47}, 777-780.
\bibitem{}
Freedman, S. J., and J. S. Clauser, 1972, "Experimental test of
local hidden-variable theories," Phys. Rev. Lett. {\bf 28},
938-941.
\bibitem{}
Gisin, N., 1991, "Bell's inequality holds for all non-product
states," Phys. Lett. A {\bf 154}, 201-202.
\bibitem{}
Gisin, N. and A. Peres, 1992, "Maximal violation of Bell's
inequality for arbitrary large spin," Phys. Lett. A {\bf 162},
15-17.
\bibitem{}
Gnedenko, B. V., 1976, {\it The Theory of Probability} (Mir
Publishers, Moscow).
\bibitem{}
Greenberger, D. M., M. Horne, and A. Zeilinger, 1989, "Going
beyond Bell's theorem" in {\it Bell's Theorem, Quantum Theory, and
Conceptions of the Universe} edited by M. Kafatos (Kluwer
Academic, Dordrecht), p. 73-76.
\bibitem{}
Greenberger, D. M., M. Horne, A. Shimony, and A. Zeilinger, 1990,
"Bell's theorem without inequalities," Am. J. Phys. {\bf 58},
1131-1143.
\bibitem{}
Horodecki, R., P. Horodecki, and M. Horodecki, 1995, "Violating
Bell inequality  by mixed spin-1/2 states: Necessary and
sufficient condition," Phys. Lett. A {\bf 200}, 340-344.
\bibitem{}
Horodecki, R. and M. Horodecki, 1996, "Information-theoretic
aspects of inseparability of mixed states," Phys. Rev. A {\bf 54},
1838-1843.
\bibitem{}
Kaszlikowski, D., P. Gnaciski, M. \.{Z}ukowski, W. Miklaszewski,
and A. Zeilinger, 2000, "Violations of Local Realism by Two
Entangled N-Dimensional Systems Are Stronger than for Two Qubits,"
Phys. Rev. Lett. {\bf 85}, 4418-4421.
\bibitem{}
Mermin, N. D., 1990, "Extreme quantum entanglement in a
superposition of macroscopically distinct states," Phys. Rev.
Lett. {\bf 65}, 1838-1841.
\bibitem{}
Mermin, N. D., 1998, "What is quantum mechanics trying to tell
us?" Am. J. Phys. {\bf 66}, 753-767.
\bibitem{}
Pan, J. W., D. Bouwmeester, H. Weinfurter, and A. Zeilinger, 2000,
"Experimental test of quantum nonlocality in three photon
Greenberger-Horne-Zeilinger entanglement," Nature {\bf 403},
515-518.
\bibitem{}
Peres A., 1999, "All the Bell inequalities,"  Found. Phys. {\bf
29}, 589-614.
\bibitem{}
Pitowsky I., 1989, {\it Quantum Probability - Quantum Logic}
(Springer, Berlin).
\bibitem{}
Pitowsky I. and K. Svozil, 2001, "Optimal tests of quantum
nonlocality", Phys. Rev. A {\bf 64}, 014102.
\bibitem{}
Popescu, S., 1995, "Bell's inequalities and density matrices:
Revealing 'hidden' nonlocality," Phys. Rev. Lett. {\bf 74},
2619-2622.
\bibitem{}
Scarani, V. and N. Gisin, 2001, "Spectral decomposition of Bell's
operators for qubits," Preprint quanth-ph/0104016 at
(http://xxx.lanl.gov).
\bibitem{}
Schr\"odinger, E., 1935a, "Die gegenw\"artige Situation in der
Quantenmechanik," Naturwissenschaften 23, 807-812; 823-828;
844-849. Translation published in {\it Proc. Am. Phil. Soc.} 124,
p. 323-338 and in {\it Quantum Theory and Measurement} edited by
J. A. Wheeler and W. H. Zurek(Princeton University Press,
Princeton), p. 152-167. A copy can be found at
(www.emr.hibu.no/lars/eng/cat).
\bibitem{}
Schr\"odinger, E., 1935b, "Discussion of probability relations
between separated systems" in {\it Proceedings of the Cambridge
Philosophical Society} 31, p. 555.
\bibitem{}
Shannon, C. E., 1948, ``A mathematical theory of communication,''
Bell Syst. Tech. J. {\bf 27}, 379. A copy can be found at \\
(http://cm.bell-labs.com/cm/ms/what/shannonday/paper.html).
\bibitem{}
von Weizs\"acker, C. F., 1975, in {\it Quantum Theory and the
Structure of Time and Space II} edited by L. Castell, M.
Drieschner and C. F. Weizs\"acker (Hanser, M\"{u}nchen).
\bibitem{}
Weihs, G., T. Jennewein, C. Simon, H. Weinfurter, and A.
Zeilinger, 1998, "Violation of Bell's inequality under strict
Einstein locality conditions," Phys. Rev. Lett. {\bf 81},
5039-5043.
\bibitem{}
Weinfurter, H. and M. \.{Z}ukowski M, 2001, "Four-Photon
Entanglement from Down-Conversion," Phys. Rev. A {\bf 64},
010102(R).
\bibitem{}
Werner, R. F. and M. M. Wolf, 2001, "All multipartite Bell
correlation inequalities for two dichotomic observables per site,"
Preprint quant-ph/0102024 at (http://xxx.lanl.gov).
\bibitem{}
Zeilinger, A., 1997, "Quantum teleportation and the non-locality
of information," Phil. Trans. Roy. Soc. Lond. {\bf 1733},
2401-2404.
\bibitem{}
Zeilinger, A., 1999, "A foundational principle for quantum
mechanics," Found. Phys. {\bf 29}, 631-643.
\bibitem{}
\.{Z}ukowski, M., 1993, "Bell Theorem Involving all Settings of
Measuring Apparata," Phys. Lett. A {\bf 177} 290-294.
\bibitem{}
\.{Z}ukowski, M. and \v{C} Brukner, 2000, "Bell's theorem for
general N-qubit states," Preprint quant-ph/0102039 at
(http://xxx.lanl.gov).
\bibitem{}
\.Zukowski, M. and D. Kaszlikowski, 1997, "Critical visibility for
N-particle Greenberger-Horne-Zeilinger correlations to violate
local realism," Phys. Rev. A {\bf 56}, R1682-R1685.


\end{references}
\end{document}